# Artificial Synapse Network on Inorganic Proton Conductor for Neuromorphic Systems Applications

Li Qiang Zhu[1,2], Chang Jin Wan[1,2], Li Qiang Guo[1], Yi Shi[1], and Qing Wan[1, 2]

The basic units in our brain are neurons and each neuron has more than 1000 synapse connections. Synapse is the basic structure for information transfer in an ever-changing manner, and short-term plasticity allows synapses to perform critical computational functions in neural circuits. Therefore the major challenge for the hardware implementation of neuromorphic computation is to develop artificial synapse. Here, in-plane oxide-based artificial synapse network coupled by proton neurotransmitters are self-assembled on glass substrates at room-temperature. A strong lateral modulation is observed due to the proton migration in the nanogranular phosphorus-doped $SiO_2$ electrolyte films. Functional roles of short-term plasticity, including paired-pulse facilitation, dynamic filtering and spatiotemporally correlated signal processing are mimicked. Such in-plane oxide-based protonic/electronic hybrid synapse network proposed here are interesting for building future neuromorphic systems.

[1] School of Electronic Science & Engineering, Nanjing University, Nanjing 210093, People's Republic of China; [2] Ningbo Institute of Materials Technology and Engineering, Chinese Academy of Sciences, Ningbo 315201, People's Republic of China. Correspondence and requests for materials should be addressed to Q.Wan (E-mail: wanqing@nimte.ac.cn)

Human beings could easily recognize various objects and make sense out of a large amount of visual information in a complex, real world environment through a complicated computation, which is more robust, plastic and fault-tolerant than any current digital computer. Synapses dominate the architecture of the brain and are responsible for the massive parallelism, structural plasticity, and robustness.[1] Synapses are the connections between the neuron circuits, and each neuron has more than 1000 synapse connections with other neurons. As a measurement of synapse efficacy in establishing the connection between pre-synaptic and post-synaptic neurons, synaptic strength or synaptic weight, can be precisely adjusted by the concentrations of ionic species (e.g., $Ca^{2+}$, $Na^+$, and $K^+$, etc) [2-4] Realization of physical devices with synaptic functions is of great importance for hardware implementation of neuromorphic computation system. In the beginning, synaptic functions were emulated by CMOS neuromorphic circuits, but such CMOS circuits consumed substantially more energy than a biological synapse, and it is hard to scale up the circuits to a size comparable with brain.[5] Recently, resistive switching memory, memristors or atomic switch have been investigated in biologically inspired neuromorphic circuits.[6-11] Important synaptic learning rules such as spike-timing dependent plasticity (STDP) and short-term memory to long-term memory transition have been demonstrated. In these nanoscale ionic/electronic hybrid two-terminal artificial synapses, ionic migrant/diffusion, formation of conducting filament and electrochemical reactions are very prevalent for mimicking the neural activities.[7, 10]

In biologic systems, the dendrites of a single neuron can contain hundreds to thousands of spines which typically receive inputs from synapses of axons.[12] The dendrite spine synapses not only provide an anatomical substrate for memory storage and synaptic transmission, but also serve to increase the number of possible contacts between neurons. Recently, three-terminal transistors with tunable channel conductivities have been proposed for synaptic functions emulation. [13-17] In such synaptic transistors, voltage pulses applied on the gate electrodes are usually regarded as the synaptic spikes or external stimuli. The channel conductance is usually regarded as synaptic weight. Then, such synaptic transistors are similar to the dendrite spine synapses in biologic systems. Y. Chen et al proposed an ionic/electronic hybrid synaptic transistor by integrating an $RbAg_4I_5$ ionic conductor layer and an ion-doped MEH-PPV conjugated polymer layer into the gate dielectrics. [16] Essential synaptic dynamic functions like excitatory/inhibitory postsynaptic current (EPSC/IPSC) and STDP were successfully mimicked.

For a conventional field-effect transistor, bottom-gate or top-gate configuration is always adopted. Figure 1a shows a schematic image of a bottom-gate transistor, where the gate electrode is buried below the gate dielectric and the semiconductor channel [18, 19]. In such structure, the gate bias is coupled to the semiconductor channel through the dielectric directly. Recently, oxide-based protonic/electronic hybrid transistors with in-plane-gate configuration were invented by our group. [20-23] As shown in figure 1b, proton conductors, such as $SiO_2$-based nanogranular films and chitosan films, were used as the gate dielectrics. Voltage applied on the in-plane-gate is coupled to the common bottom conductive layer first, and then it can be coupled to the channel layer, ie, the gate bias will be coupled to the channel through two gate capacitors in series. In this communication, a new type of transistor is proposed based on the lateral coupling effect by proton conducting electrolyte film. As shown in figure 1c, no bottom conductive layer is needed, and the gate voltage can be directly coupled to the semiconductor channel laterally by only one lateral electric-double-layer (EDL) capacitor. The indium-zinc-oxide (IZO)-based protonic/electronic hybrid transistors were self-assembled on phosphorus-doped nanogranular $SiO_2$ proton conductive films at room temperature. The detailed process description can be found in Supplementary Information. Thanks to such lateral proton-related coupling effect, artificial synaptic transistor network can be fabricated, as shown in figure 1d. Dynamic logic, including paired-pulse facilitation, dynamic filter for information transmission and spatiotemporally correlated signal processing, are successfully mimicked. Synaptic transistor network proposed here is similar to the dendrite spine synapses in biologic system, and is interesting for building neuromorphic systems.

**Results**

Figure 2a shows a cross-sectional scanning electron microscopy (SEM) image of the P-doped $SiO_2$ electrolyte film on a Si (100) substrate. Microstructure with nanopores is observed. Inset in figure 2a illustrates transmission electron microscope (TEM) images of the P-doped nanogranular $SiO_2$ film deposited on TEM Cu grid. Nanogranular $SiO_2$ with the size in tens of nanometer is observed. The size of the gap among the $SiO_2$ nanoparticles is ~2.0 nm to 5.0 nm. Therefore, high-density nanopores are formed in the electrolyte films. Such nanopores provide effective channels for proton migration. Electrical properties of the P-doped nanogranular $SiO_2$ films were characterized by impedance spectroscopy with in-plane IZO electrode structure, as shown by the inset in figure 2b. Frequency dependent specific capacitances in the frequency range from 1.0 Hz to 10.0 MHz are shown in figure 2b. A maximum specific capacitance of ~3.0 $\mu F/cm^2$ is measured at 1.0 Hz, which is due to the electric-double-layer (EDL) formation at the $SiO_2$ electrolyte/IZO electrode interface. Detail discussion can be found in Supplementary Information. Proton conductivity of the nanogranular P-doped $SiO_2$ film was estimated to be ~$10^{-4}$ S/cm at room temperature. [23] High proton conductivity is due to the sequence of proton hopping between hydroxyl groups and water molecules under applied electric field. [23-25] Accumulation of protons at the interface of $SiO_2$ electrolyte/IZO electrode will result in large EDL capacitance and strong lateral electrostatic coupling effect, which is meaningful for potential applications in artificial synaptic network. [26-28]

The conductance of the self-assembled IZO channel between two patterned IZO

films (source and drain electrodes) can be effectively modulated by another in-plane IZO pattern (gate electrode). So the fabricated device is a lateral-coupled IZO-based thin-film transistor, as shown in the inset of figure 2c. Transfer curve ($I_{ds}$ vs $V_{gs}$) was measured with a constant $V_{ds}$ of 1.0 V for device with a channel thickness of ~20 nm. The gate voltage was swept from -1.0 V to 1.0 V and then back. A clear anticlockwise hysteresis loop was observed, which is likely due to the existence of mobile protons in the nanogranular P-doped $SiO_2$ electrolyte. [29, 30] Figure 2d shows a typical output curve of the in-plane coupled transistor. The drain current on/off ratio is estimated to be ~$10^6$. At low $V_{ds}$, the drain current increases linearly with drain voltage, indicating that the device has a good ohmic contact. At higher $V_{ds}$, the drain current gradually approaches to a saturated value. Such results indicate that the measured device is a typical lateral-coupled field-effect transistor.

In nervous system, synapse permits a neuron to pass an electrical or chemical signal to another neuron [31]. Figure 3a illustrates a simplified biological dendrite spine synapse. It could establish a relationship between pre-synapse and post-synapse, therefore could provide an anatomical substrate for synaptic transmission. For the lateral-coupled IZO transistor proposed here, IZO channel conductance could be modulated significantly by proton migration within nanogranular $SiO_2$ layer under external electrical field (shown in Fig. 2c). Taking the in-plane gate as pre-synapse and IZO channel with IZO source/drain as post-synapse, an IZO synaptic transistor is proposed, as shown in figure 3b. The synaptic transistor could mimic spike triggered temporal response of biologic synapse. A presynaptic spike applied on the in-plane-gate will trigger an excitatory post-synaptic current (EPSC) in IZO channel. Figure 3c shows typical EPSC of a synaptic transistor with a channel thickness of 12 nm triggered by a presynaptic spike (0.3V, 10 ms). A constant Vds of 0.5V is used for channel current measurement. The EPSC reaches a peak value of ~13 nA at the end of the spike and gradually decays back to the resting current (~5.0 nA) in ~500 ms. Such EPSC behavior is quite similar to an EPSC process in a biological excitatory synapse.[32] The energy dissipation of single spike event is estimated to be ~45 pJ, which is lower than that of the artificial synapse based on conventional CMOS circuit [33] and is comparable to the energy dissipation of the reported hard-ware based artificial synapse [8, 9, 15] At present, the channel width of our synaptic transistors is large (~150 μm), and it can be scaled down to sub-micrometer scale by photolithography method. Such scaling can further reduce the energy dissipation. The scaling can potentially compromise the voltage level, noise margin and low-frequency noise, but for synaptic information process, the arithmetic is based on timing-dependent spiking. So for a neuron network, noise is not a big problem as in the case of traditional digital IC chip. Low-voltage and low energy dissipation down to 1.0 pJ/spike is the aim for neuromorphic system application. Moreover, the energy dissipation per spike can also be reduced to ~1.0 pJ level or even sub-pJ level if the spike duration can be reduced to sub-ms level. Here, the spike duration dependent excitatory post-synaptic currents (EPSCs) are also studied, as shown in figure 3d. The measured EPSC peak values increase from ~13 nA to ~29 nA for spike duration ranges from 10 ms to 2000 ms, respectively. Interestingly, the EPSC peak value increases almost linearly within a small spike duration range below ~300 ms, but it will get saturated when the spike duration is above ~300ms.

The migration of protons within the nanogranular $SiO_2$ film plays an important role for EPSC triggering. When a positive presynaptic spike is applied on the lateral IZO in-plane gate, protons inside the nanogranular $SiO_2$ films would migrate and accumulate at the $SiO_2$/IZO channel interface, and a postsynaptic current cane be measured when a small drain voltage is applied. More protons will migrate to the $SiO_2$ electrolyte/IZO channel interface when the spike duration time is increased. So, the electron density in the IZO channel increases, and an increased EPSC can be obtained. However, there are only limited activated protons within the nanogranular $SiO_2$ matrix. Therefore, the measured EPSC in the IZO channel will become saturated at a certain spike duration time.

When the spike is finished, the protons would gradually drift back to their equilibrium positions in the SiO$_2$ matrix. Correspondingly, the measured current in the self-assembled IZO channel will decrease gradually back to the resting current. Such proton migration induced EPSC process can be explained with the stretched exponential model and the decay of EPSC could be fitted by a stretched exponential function (SEF): [8, 34]

$$I = (I_0 - I_\infty)\exp[-(\frac{t-t_0}{\tau})^\beta] + I_\infty$$

where $\tau$ is the retention time, $t_0$ is the time when the pre-synaptic spike finished, $I_0$ is the triggered EPSC at the end of the pre-synaptic spike, $I_\infty$ is the final value of the decay current and $\beta$ is the stretch index ranging between 0 and 1. A good fitting curve of the decay of EPSC triggered by pre-synaptic spike (0.3V, 10 ms) is obtained, as shown in figure 3e. $\tau$ is estimated to be ~25 ms, which means that the feature time of the protons migration is ~25 ms.

In order to investigate the influence of voltage applied on the second in-plane-gate to the energy dissipation per spike of the synaptic transistor, synaptic transistor with dual in-plane-gate configuration is used to measure the spike triggered EPSC. A presynaptic spike 0.3 V, 10 ms) is applied on one in-plane IZO gate, and a constant $V_{ds}$ of 0.5V is used to measure the EPSC. Different voltages are applied on the second in-plane IZO gate electrode in order to investigate their influence to the energy dissipation per spike. Figure 3f shows the energy dissipation per spike modulated by the applied voltage on the second in-plane gate. When there is no bias on the second in-plane-gate (G), the energy dissipation per spike for the synaptic transistors with a channel thickness of 20 nm is ~1.4 nJ. When the voltage applied on the second in-plane-gate (G) is changed from 0 V to -0.7 V, the energy dissipation reduces progressively from ~1.4 nJ to ~15 pJ. At the same time, EPSC current is also depended on the distance between the pre-synapse and the post-synapse. Increasing the distance will reduce the EPSC value. The detailed description can be found in Supplementary Information.

In neuronal networks, temporal dynamics is a general feature of synaptic transmission. As one of many time-dependent neuronal properties, short-term synaptic plasticity (STP) could provide a memory of the recent stimulus history in a network.[35] Paired-pulse facilitation (PPF) is a common form of short-term synaptic plasticity in biological synapses reflecting a responsivity for the second pulse when first pulse and the second pulse are close enough.[36] The first spike will cause a small post-synaptic response, while the second spike will cause the increased post-synaptic response. As a result, greater transmitter release occurs, ie, the higher PPF. Such PPF is essential to decode temporal information in biological system.[37] IZO synaptic transistor proposed here could mimic the temporally correlated spikes and generate temporal analog logic. Two pre-synaptic spikes (0.3 V, 10 ms) are applied on pre-synapse (IZO gate) in close succession. The inter-spike interval ($\Delta t_{pre}$) ranges between 30 ms and 1500 ms. The responses of the post-synapse are measured in terms of excitatory postsynaptic current (EPSC) with a constant $V_{ds}$ of 0.5 V. Figure 4a shows the EPSC triggered by the two successive pre-synaptic spikes with $\Delta t_{pre}$=200 ms. The EPSC triggered by the second pre-synaptic spike is larger than that by the first pre-synaptic spike. The paired-pulse facilitation index, defined as the ratio of the amplitudes between the second EPSC (A2) and the first EPSC (A1), is plotted as a function of $\Delta t_{pre}$ in figure 4b. A maximum PPF index value of ~180% is obtained at $\Delta t_{pre}$=30 ms. When $\Delta t_{pre}$ increases, PPF index value decreases gradually. At the end of the first spike, protons accumulated at the $SiO_2$/IZO channel interface will drift gradually back to their equilibrium position due to the concentration gradient. If the second spike is applied after the first spike with a small inter-spike interval, protons triggered by the first spike still partially reside near the IZO channel. Thus, protons triggered by the second spike are augmented, which induces the PPF and short-term memory in IZO channel. A higher PPF index value can be obtained when the applied paired spikes have a shorter $\Delta t_{pre}$.

In nervous system, synaptic efficacy can increase (synaptic facilitation) or decrease

(synaptic depression) markedly within milliseconds after the onset of specific temporal patterns of activity.[38] Due to the short-term synaptic depression or facilitation, synapses could also act as dynamic filters for information transmission depending on the signal frequencies. [39] The short-term synaptic depression contributes to low-pass temporal filtering and the short-term synaptic facilitation contributes to high-pass temporal filtering. Since a higher PPF index value is obtained with the shorter $\Delta t_{pre}$ in our IZO synaptic transistor, a high-pass filter could be realized. Figure 4c shows the EPSC responses of our device to the stimulus train with different frequencies. The stimulus train at each frequency is consisted of 10 stimulus spikes (0.3V, 10ms). $V_{ds}$=0.5 V is applied to the source and drain electrodes for EPSCs measurements. For low frequency at 1Hz, the peak value of the EPSC keeps at ~340 nA even after 10 spikes when the frequency of the stimulus train is 1.0 Hz. When the frequency of the stimulus train is increased, the peak values of the measured EPSCs increase obviously. Figure 4d shows the frequency dependent gain defined as the ratio of the amplitudes between the tenth EPSC (A10) and the first EPSC (A1). The gain increases from ~1.0 to ~6.6 when the stimulus frequency changes from 1.0 Hz to 50 Hz, which indicates that the synaptic transistor can act as the dynamic high-pass filter for information transmission. The high-pass temporal filtering mimicked here is meaningful for neuromorphic computations or artificial neural network.

The spatiotemporally correlated stimuli from different neurons would trigger a post-synaptic neuron to establish dynamic logic. In order to simulate the spatiotemporal dynamic logic, lateral coupled synaptic transistor with two in-plane gates is proposed, as shown in figure 5a. The responses of the post-synapse are measured with a constant $V_{ds}$ of 0.5V). Pre-synaptic spike (0.5V, 20 ms) on pre-synapse 1 will trigger an EPSC 1 with an amplitude of ~30 nA and pre-synaptic spike (1.0 V, 20 ms) on pre-synapse 2 will trigger an EPSC 2 with an amplitude of ~50 nA, as shown in figure 5b. When the two pre-synaptic spikes with an inter-spike interval ($\Delta t_{pre2-pre1}$) are applied on pre-synapse 1 and pre-synapse 2 respectively, the two EPSCs will be summed in the post-synapse, which is a dynamic analog function of time and $\Delta t_{pre2-pre1}$. Figure 5c illustrates the amplitude of the EPSC at $\Delta t=0$ (when the pre-synaptic spike applied on the Synapse 1 finished) as a function of the inter-spike interval $\Delta t_{pre2-pre1}$. When $\Delta t_{pre2-pre1}=0$, the EPSC1 and EPSC2 are triggered simultaneously, therefore the amplitude of the EPSC in the post-synapse reaches the maximum value of ~110 nA. The summed amplitude (~110 nA) is larger than the linear sum of EPSC 1 (~30 nA) and EPSC 2 (~50 nA). The results show the spatiotemporal properties of supralinear amplification in EPSC summation on post-synapse. Such supralinear amplification is similar to that observed in hippocampal CA1 pyramidal neurons.[40] When $|\Delta t_{pre2-pre1}|$ increases, the amplitude of the EPSC decreases asymmetrically. When the IZO pre-synapse 1 is triggered earlier than the IZO pre-synapse 2 ($\Delta t_{pre2-pre1}>0$), the amplitude of EPSC at $\Delta t=0$ is equal to the peak amplitude of EPSC1. Whereas IZO synapse 1 is triggered later than IZO Synapse 2 ($\Delta t_{pre2-pre1}<0$), the amplitude of EPSC at t = 0 is the summed current of EPSC1 from the IZO pre-synapse 1 and EPSC2 from the IZO pre-synapse 2. With the decrease of the $\Delta t_{pre2-pre1}$, the increase in EPSC amplitude at t = 0 by EPSC2 from the IZO pre-synapse 2 is getting less significant, ie, the effects of the pre-synaptic spike on IZO pre-synapse 2 are getting less important.

**Discussion**

Here we should point out that the number of the in-plane gate is not limited to be two, and multiple in-plane gates can be deposited on the P-doped nanogranular $SiO_2$ films through one mask deposition process, as shown in figure 6a. When these multiple in-plane IZO gates are used as the presynaptic input terminals and one self-assembled IZO channel with source/drain electrodes is used as the postsynaptic output terminals, a simple artificial neural network is obtained. Thanks to the strong lateral in-plane protonic/electronic coupling effect, no intentional hard-wire connections are needed. Figure 6b illustrates the simplified scheme for the artificial neural network with multiple presynapses. Proton migration triggered by the gate pulse results in the establishments of the inter-connections between the in-plane gates and the self-assembled IZO channel. Such process is analogue to the spike modulated movement of the neurotransmitters, which changes the synapse efficacy in establishing the relationship between pre-synapses and pos-tsynapse.[4] In such artificial neural network with multiple presynapses, the synaptic weight of the self-assembled IZO channel is spatiotemporally dependent on the spikes applied on the multiple presynaptic inputs, which are highly desirable for neuromorphic computation.

In summary, lateral in-plane-gate IZO-based synaptic transistor network was fabricated on P-doped $SiO_2$ electrolyte film. A strong lateral electrostatic coupling effect was observed due to proton migration in the nanogranular phosphorus-doped $SiO_2$ electrolyte film. Paired-pulse facilitation, dynamic filter function and spatiotemporal signal processing are successfully mimicked. Such oxide-based synapse network proposed here are interesting for building future neuromorphic systems.

**Methods**

**Fabrication of lateral coupled IZO-based synaptic transistor.** Lateral coupled IZO-based synaptic transistors were fabricated on glass substrate at room temperature. First, phosphorus (P)-doped nanogranular $SiO_2$ films were deposited on glass substrates by plasma enhanced chemical-vapor deposition (PECVD) using $SiH_4/PH_3$ mixture (95% $SiH_4$+5% $PH_3$) and $O_2$ as the reactive gases and Ar as plasma enhancement gas. Ratio frequency (RF) power is 100 W and working pressure is ~30 Pa. The flow rates of the $SiH_4/PH_3$ mixture (95% $SiH_4$+5% $PH_3$), $O_2$ and Ar are 10 sccm, 60 sccm and 60 sccm, respectively. The thickness of the P-doped nanogranular $SiO_2$ films is ~820 nm. Then, IZO patterns were deposited on the P-doped nanogranular $SiO_2$ films by radio-frequency (RF) magnetron sputtering IZO target in pure Ar ambient with a nickel shadow mask, as shown in figure 1c and figure S1. Ratio frequency (RF) power, Ar flow rate and chamber pressure are 100 W, 14 sccm and 0.5 Pa, respectively. When the distance between the patterns is 80μm, a thin IZO channel can be self-assembled between the IZO source and drain electrodes due to the reflection of IZO nanoparticles at the mask edge and dimensional extention by IZO nanoparticles with a low incident angle. The thickness of the self-assembled IZO channel layer can be controlled by the distance between metal mask and the substrate. When the distance between the patterns is as large as 300μm, no connected channel layer can be obtained and the isolated IZO patterns are used as the in-plane gates.

**Microstructure characterizations of the P-doped $SiO_2$ film.** For microstructure characterization, P-doped $SiO_2$ films were also deposited on a polished Si (100) wafers and Cu TEM grids. Field-emission scanning electron microscopy (FE-SEM) and transmission electron microscopy (TEM) were used for structural characterization of the nanogranular P-doped $SiO_2$ films.

**Electrical characterizations of the P-doped SiO$_2$ film and lateral-coupled IZO-based synaptic transistors.** Proton conductivity and frequency dependent capacitances of the P-doped SiO$_2$ electrolyte films were characterized by Solartron 1260A Impedance Analyzer in air ambient with a relative humidity of ~50%. Electrical performance of the transistors and the simulation of synaptic functions were measured by the semiconductor parameters characterization system (Keithley 4200 SCS) under air relative humidity of ~50%. Presynaptic spikes are applied on the in-plane gate electrode, and postsynaptic output is measured by applying a small reading drain voltage on the IZO source/drain electrodes.

**Figure Legends**

**Figure 1 | Comparison between the previous thin-films transistor and the synaptic transistor in this work.** (a) Conventional bottom gate thin-films transistors. (b) In-plane-gate oxide thin-films transistor with a bottom conductive layer. (c) Schematic image of a self-assembled IZO-based transistor fabricated by one-mask method in this work. (d) Schematic image of the IZO-based protonic/electronic hybrid synaptic transistor. In-plane gates are taken as pre-synapses, while the IZO patterns with self-assembled IZO channel is taken as post-synapse.

**Figure 2 | Microstructure and lateral specific capacitance of the P-doped nanogranular $SiO_2$ electrolyte film and electrical performances for the lateral-gated IZO transistor.** (a) Cross-sectional SEM images of the P-doped nanogranular $SiO_2$ film. Inset: TEM image of the P-doped nanogranular $SiO_2$ films on TEM Cu grid. (b) Frequency-dependent specific capacitance of the P-doped nanogranular $SiO_2$ electrolyte film. Inset: in-plane structure for impedance spectroscopy characterization. (c) Transfer curve ($I_{ds}$ vs $V_{gs}$ curve) measured with a constant $V_{ds}$ of 1 V. Inset: schematic image of the lateral coupled transistor. (d) Output characteristics of the lateral coupled transistor.

**Figure 3 | Comparison between the biological synapse and the IZO artificial synapse and the excitatory post-synaptic current (EPSC) triggered by pre-synaptic spikes.** (a) A simplified biological dendrite spine synapse. Pre-synaptic spike governs the movements of neurotransmitters, modulating the synapse efficacy in establishing the relationship between the pre-synapse and the post-synapse. (b) Schematic diagram showing the IZO-based protonic/electronic hybrid synaptic transistor and the distributions of the mobile protons in nanogranular $SiO_2$ film. (c) Excitatory

post-synaptic current (EPSC) triggered by pre-synaptic spike (0.3 V, 10 ms). The EPSC is measured with a constant $V_{ds}$ of 0.5 V. **(d)** Spike duration dependent EPSC peak values with a spike amplitude of 0.3 V. **(e)** SEF fitting for the decay of EPSC triggered by pre-synaptic spike (0.3V, 10 ms). **(f)** Energy dissipation per-spike modulated by the bias on the second in-plane gate. EPSC is triggered by pre-synaptic spike of 0.3V in amplitude and 10ms in duration and measured by a constant $V_{ds}$ of 0.5 V.

**Figure 4 | Paired-pulse facilitation and dynamic filter behaviors of the lateral gated IZO synaptic transistor.** **(a)** A pair of pre-synaptic spikes and the triggered EPSC under an inter-spike interval time of 200 ms. A1 and A2 represent the amplitudes of the first and second EPSCs, respectively. **(b)** PPF index, defined as the ratio of A2/A1, plotted as a function of inter-spike interval, $\Delta t_{pre}$, between the two consecutive spikes. **(c)** EPSCs recorded in response to stimulus train with different frequencies (1.0 Hz, 5.0 Hz, 10 Hz, 20 Hz, 25 Hz, 33Hz and 50 Hz). The stimulus train at each frequency is consisted of 10 stimulus spikes (0.5V, 10ms). **(d)** EPSCs amplitude gain (A10/A1) plotted as a function of pre-synaptic spike frequency.

**Figure 5 | Spatiotemporally correlated pre-synaptic spikes from different synapses trigger a post-synapse to establish dynamic logic.** **(a)** An in-plane synaptic device for simulating the spatiotemporal dynamic logic. Pre-synaptic spikes are applied on the in-plane IZO gates working as the pre-synapses. While a constant $V_{ds}$ 0.5 V to measure the post-synaptic current. **(b)** Post-synaptic current triggered by pre-synaptic spike 1 (0.5

V, 20 ms) and pre-synaptic spike 2 (1.0 V, 20 ms), respectively **(c)** The amplitude of the EPSC at Δt=0 (when the pre-synaptic spike applied on the pre-synapse 1 finished) is plotted as a function of $\Delta t_{pre2-pre1}$ between the two pre-synaptic spikes.

**Figure 6 | Artificial neural network based on lateral-coupled IZO synaptic transistor. (a)** A schematic illustration of an artificial neural network when taking multiple in-plane IZO gates as pre-synapses and IZO pattern with self-assembled IZO channel as post-synapse. **(b)** Simplified scheme for the proposed artificial neural network with multiple pre-synapses.

**Figures**

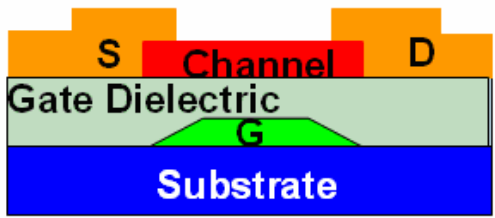 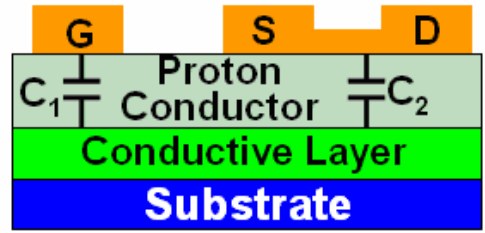
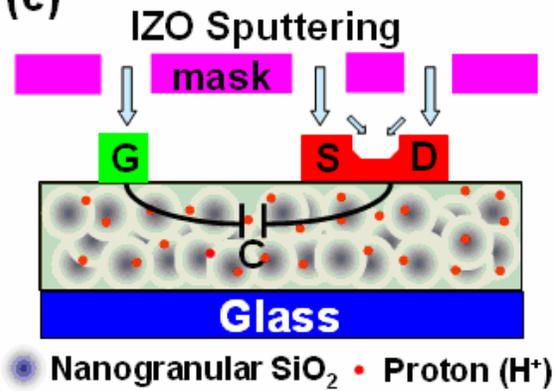 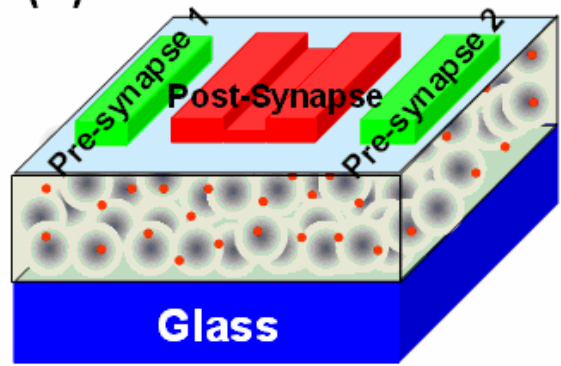

Figure 1

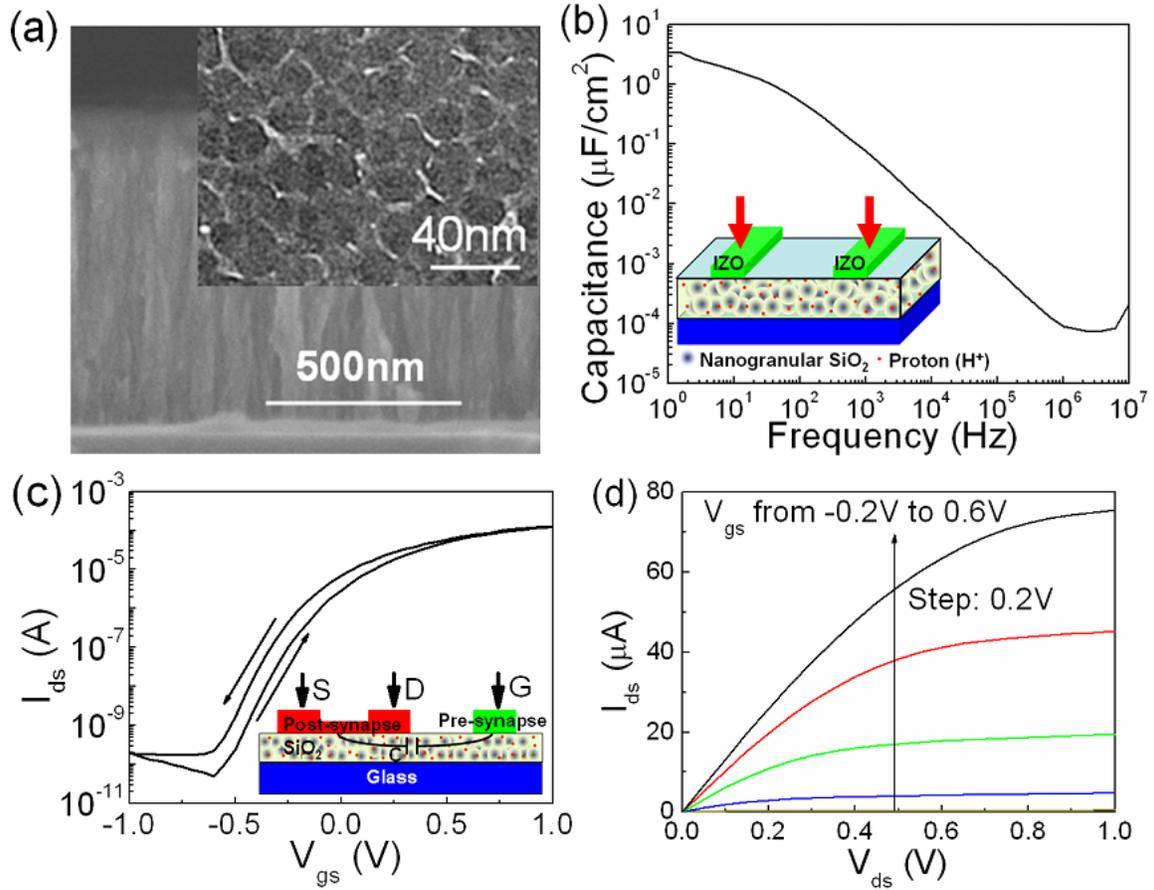

**Figure 2**

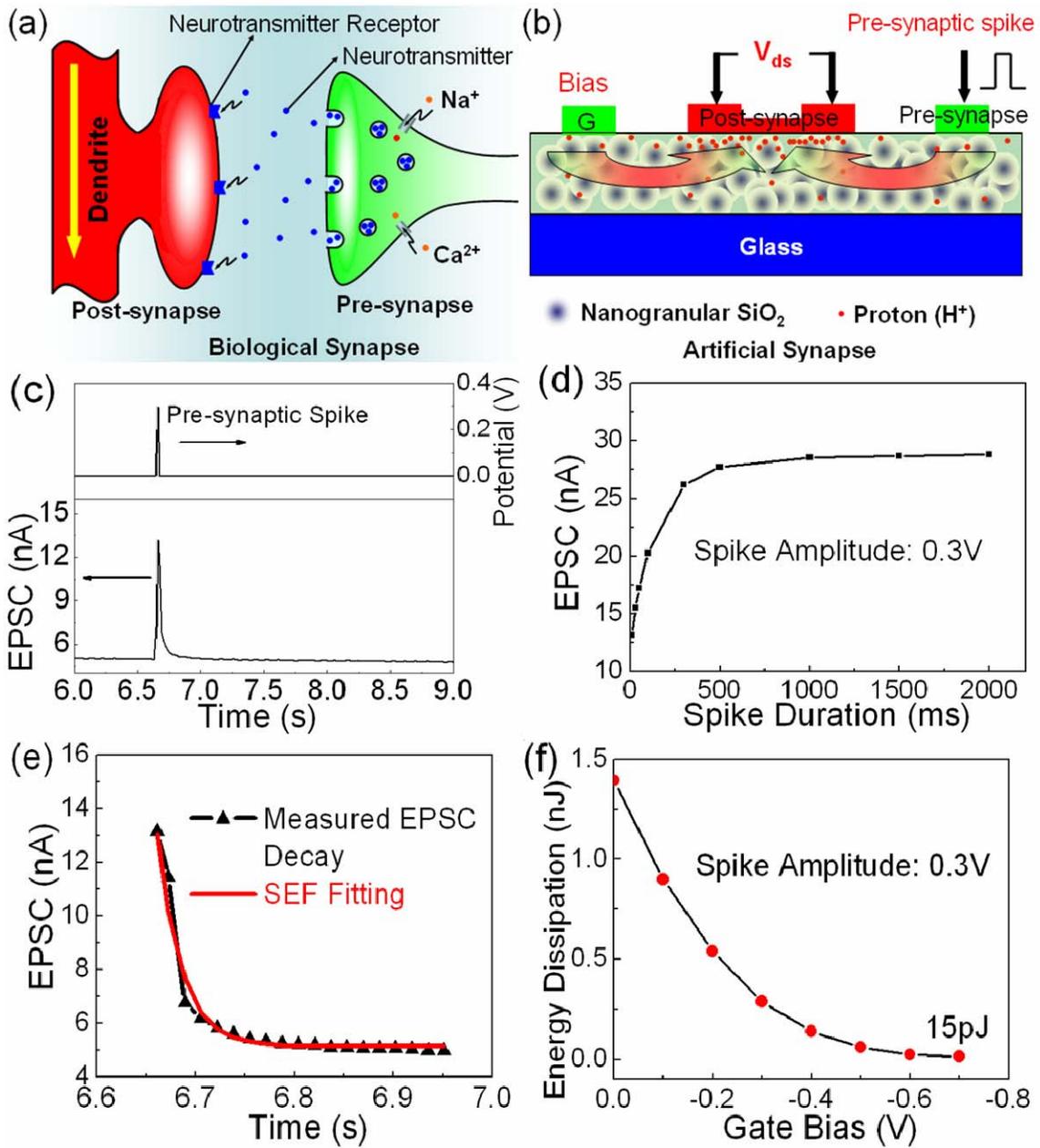

**Figure 3**

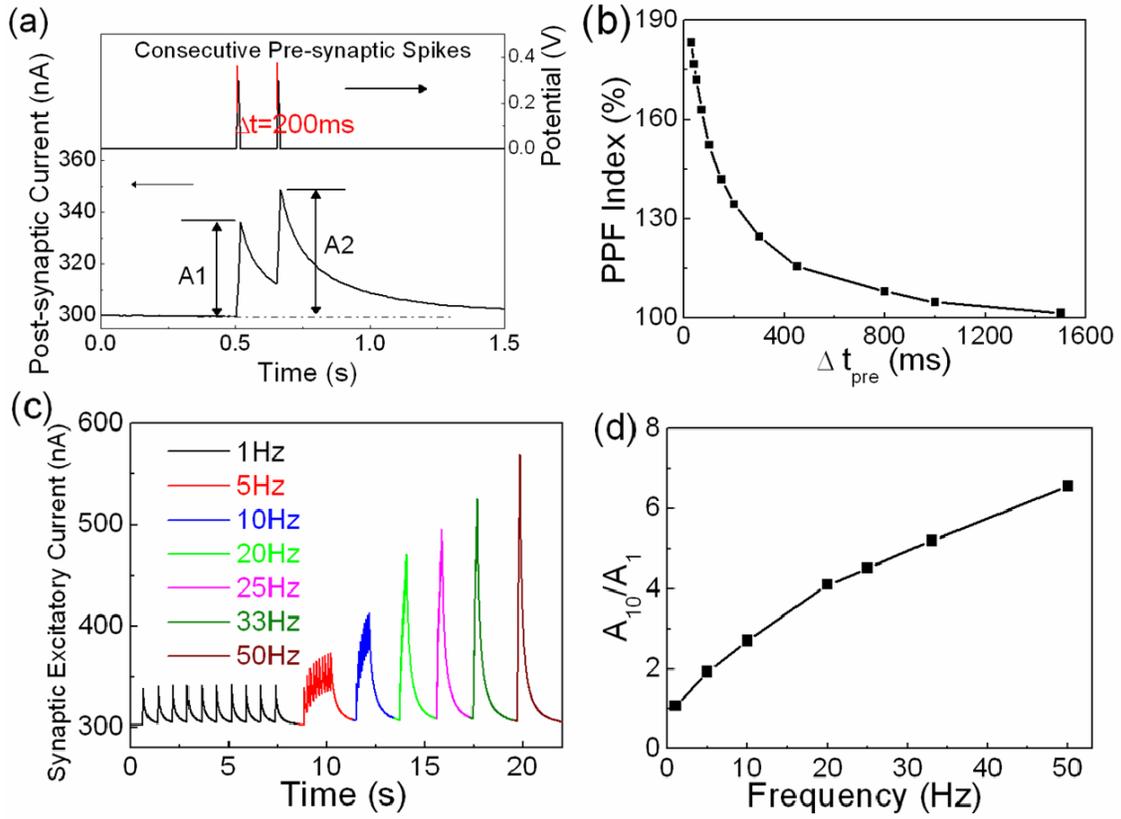

**Figure 4**

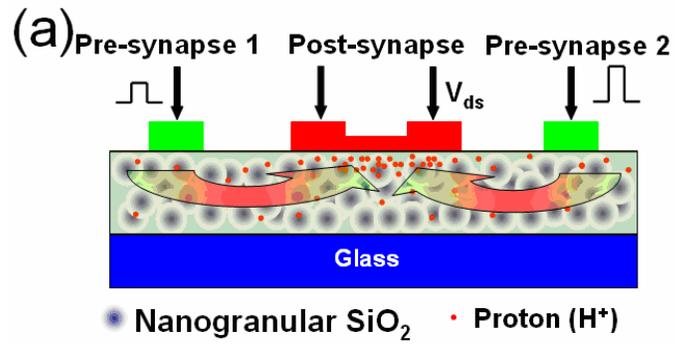
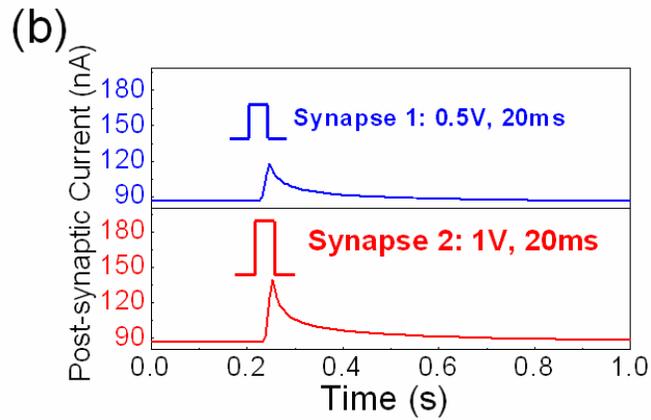
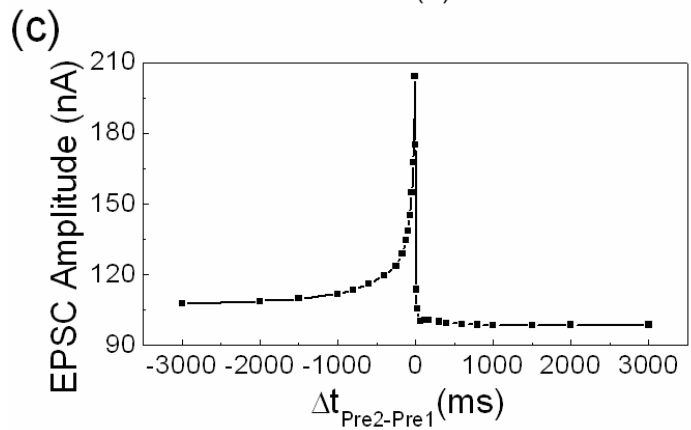

**Figure 5**

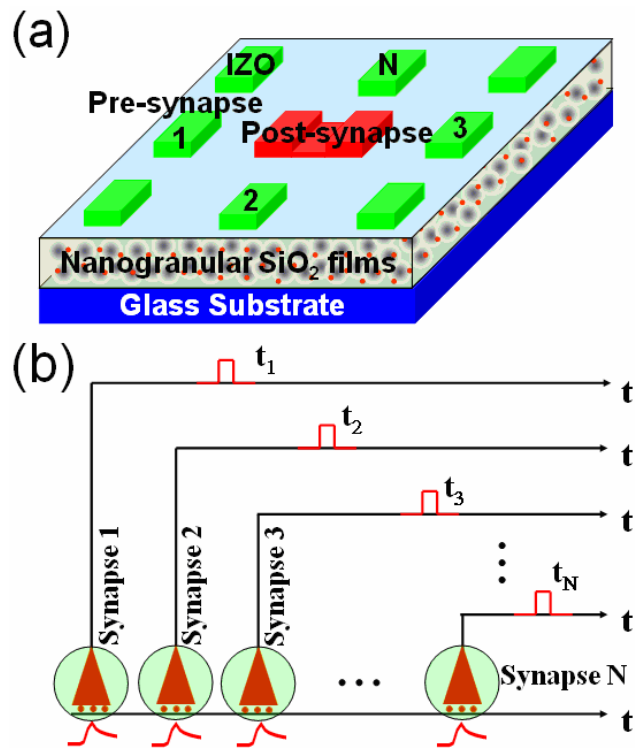

Figure 6

**Acknowledgements**

This work was supported by the National Program on Key Basic Research Project (2012CB933004), the National Natural Science Foundation of China (11174300, 11104288).

**Author contributions**

The manuscript was prepared by L.Q.Z and Q. W. Device fabrication was fabricated by L.Q.G. and L. Q. Z. Measurements were performed by L.Q.Z. and C.J.W. The project was guided by Q.W. and Y.S.

**Additional information**

**Supplementary Information** accompanies this paper on http://www.nature.com/naturecommunication

# References

1. Drachman, D.A. Do we have brain to spare? Neurology **64**, 2004-2005 (2005).

2. Bi, G. Q.; Poo, M. M. Synaptic Modifications in Cultured Hippocampal Neurons: Dependence on Spike Timing, Synaptic Strength, and Postsynaptic Cell Type. J. Neurosci. **2**, 10464–10472 (1998);

3. Zucker, R. S.; Regehr, W. G. Short-Term Synaptic Plasticity. Annu. Rev. Physiol. **64**, 355–405 (2002).

4. Voglis, G.; Tavernarakis, N. The role of synaptic ion channels in synaptic plasticity. EMBO Reports **7**, 1104-1110 (2006)

5. IBM unveils a new brain simulator, IEEE Spectrum, Nov 2009.

6. Ohno, T.; Hasegawa, T.; Tsuruoka, T.; Terabe, K.; Gimzewski, J.K.; Aono, M. Short-term plasticity and long-term potentiation mimicked in single inorganic synapses. Nat.Mater. **10**, 591-595 (2011);

7. Jo, S.H.; Chang, T.; Ebong, I.; Bhadviya, B.B.; Mazumder, P.; Lu, W. Nanoscale Memristor Device as Synapse in Neuromorphic Systems. Nano lett., **10**, 1297-1301 (2010).

8. Chang, T.; Jo, S-H.; Lu, W. Short-term memory to long-term memory transition in a nanoscale memristor. ACS Nano., **5**, 7669-7676 (2011).

9. Yu, S.; Gao, B.; Fang, Z.; Yu, H.; Kang, J.; Philp Wong, H.-S. A Low Energy Oxide-Based Electronic Synaptic Device for Neuromorphic Visual Systems with Tolerance to Device Variation. Adv.Mater., **25**, 1774-1779 (2013),

10. Wang, Z.Q.; Xu, H.Y.; Li, X.H.; Yu, H.; liu, Y.C.; Zhu, X.J. Synaptic Learning and Memory Functions Achieved Using Oxygen Ion Migration/Diffusion in an Amorphous InGaZnO Memristor. Adv.Funct.Mater., **22**, 2759-2765 (2012).

11. Yu, S.; Wu, Y.; Jeyasingh, R.; Kuzum, D.; Philp Wong, H.-S. An Electronic Synapse Device Based on Metal Oxide Resistive Switching Memory for Neuromorphic Computation. IEEE Transactions on Electron Devices, **58**, 2729-2737 (2011)

12. Calabrese, B.; Wilson, M.S.; Halpain, S. Development and regulation of dendritic spine synapses. Physiology **21**, 38-47 (2006).

13. Brink, S.; Koziol, S.; Ramakrishnan, S.; Hasler, P. A biophysically based dendrite model using programmable floating-gate devices., Proceedings of the 2008 IEEE International Symposium on Circuits and Systems (ISCAS), 432-435.

14. Ramakrishnan, S., Hasler, P.; Gordon, C. Floating gate synapses with spike time dependent plasticity., Proceedings of the 2010 IEEE International Symposium on Circuits and Systems (ISCAS), 369-372.

15. Kim, K.; Chen, C-L.; Truong, Q.; Shen, A.M.; Chen, Y. A Carbon Nanotube Synapse with Dynamic Logic and Learning. Adv.Mater., **25**, 1693-1698 (2013)

16. Lai, Q.X.; Zhang, L.; Li, Z.; Stickle, W.F.; Stanley Williams, R.; Chen, Y. Ionic/Electronic Hybrid Materials Integrated in a Synaptic Transistor with Signal Processing and Learning Functions. Adv.Mater., **22**, 2448-2453 (2010)

17. Alibart, F.; Pleutin, S.; Bichler, O.; Gamrat, C.; Serrano-Gotarredona, T.; Linares-Barranco, B.; Vuillaume, D. A Memristive Nanoparticle/Organic Hybrid Synapstor for Neuroinspired Computing Adv.Funct.Mater, **22**, 609-616 (2012).

18. Liu, S-E.; Yu, M-J.; Lin, C-Y.; Ho, G-T.; Cheng, C-C.; Lai, C-M.; Lin, C-J.; King, Y-C.; Yeh, Y-H. Influence of Passivation Layers on Characteristics of a-InGaZnO Thin-Film Transistors. IEEE Electron Device Lett., **32**, 161-163 (2011);

19. Hsieh, T-Y.; Chang, T-C.; Chen, T-C.; Chen, Y-C.; Chen, Y-T.; Liao, P-Y.; Chu, A-K.; Tsai, W-W.; Chiang, W-J.; Yan, J-Y. Application of in-cell touch sensor using photo-leakage current in dual gate a-InGaZnO thin-film transistors. Appl.Phys.Lett., **101**, 212104 (2012)

20. Jiang, J.; Sun, J.; Zhu, L.Q.; Wu, G.D.; Wan, Q. Dual in-plane-gate oxide-based thin-film transistors with tunable threshold voltage. Appl.Phys.Lett., **99**, 113504 (2011)

21. Dou, W.; Zhu, L.Q.; Jiang, J.; Wan, Q. Flexible protonic/electronic coupled neuron

transistors self-assembled on paper substrates for logic applications. Appl.Phys.Lett., **102**, 093509 (2013)

22. Zhu, L.Q.; Wu, G.D.; Zhou, J.M.; Dou, W.; Zhang, H.L.; Wan, Q. Laser directly written junctionless in-plane-gate neuron thin film transistors with AND logic function Appl.Phys.Lett., **102**, 043501 (2013)

23. Zhu, L.Q.; Sun, J.; Wu, G.D.; Zhang, H.L.; Wan, Q.; Self-assembled dual in-plane gate thin-film transistors gated by nanogranular $SiO_2$ proton conductors for logic applications. Nanoscale, **5**, 1980-1985 (2013).

24. Jin, Y. G.; Qiao, S. Z.; Costa, J. C. D. da.; Wood, B. J.; Ladewig, B. P.; Lu. G. Q. Hydrolytically Stable Phosphorylated Hybrid Silicas for Proton Conduction Adv. Funct. Mater., **17**, 3304-3311 (2007).

25. Zhong, C.; Deng, Y.; Roudsari, A. F.; Kapetanovic, A.; Anantram, M.P.; Rolandi, M. A polysaccharide bioprotonic field-effect transistor Nat. Commun., **2**, 476 (2011)

26. Dou, W.; Zhu, L.Q.; Jiang, J.; Wan, Q. Flexible Dual-Gate Oxide TFTs Gated by Chitosan Film on Paper Substrates IEEE Electron Device Letters, **34**, 259-261 (2013).

27. Zhu, L.Q.; Wu, G.D.; Zhou, J.M.; Zhang, H.L.; Wan, Q.Transparent In-Plane-Gate Junctionless Oxide-Based TFTs Directly Written by Laser Scribing IEEE Electron Device Letters, **33**, 1723-1725 (2012).

28. Kim, S.H.; Hong, K.; Xie, W.; Lee, KH.; Zhang, S.; Lodge, T.P.; Frisbie, C.D. Electrolyte-gated transistors for organic and printed electronics, Adv. Mater. **25**, 1822-1846 (2013).

29. Yoon, J.; Hong, W. K.; Jo, M.; Jo, G.; Choe, M.; Park, W.; Sohn, J. I.; Nedic, S.; Hwang, H.; Welland, M. E.; Lee, T. Nonvolatile memory functionality of ZnO nanowire transistors controlled by mobile protons, *ACS Nano*. **5,** 558-564, (2010);

30. Yuan, H. T.; Shimotani, H.; Tsukazaki, A.; Ohtomo, A.; Kawasaki, M.; Iwasa. Y. Hydrogenation-induced surface polarity recognition and proton memory behavior at protic-ionic-liquid/oxide electric-double-layer interfaces, *J. Am. Chem.* Soc. **132,**

6672-6678 (2010).

31. Schacter, Daniel L (2011). Psychology Ed 2. 41 Madison Avenue New York, NY: 10010 Worth Publishers. p. 80.

32. Smith, A.J.; Owens, S.; Forsythe, I.D. Characterisation of inhibitory and excitatory postsynaptic currents of the rat medial superior olive. J.Physiology, **529**, 681-698 (2000)

33. Indiveri, G; Chicca, E.; Douglas, R. A VLSI array of low-power spiking neurons and bistable synapses with spike-timing dependent plasticity. IEEE Trans. Neural Networks **17**, 211–221 (2006)

34. Sturman, B.; Podivilov, E.; Gorkunov, M. Origin of stretched exponential relaxation for hopping-transport models, Physical Review Letters. **91,** 176602 (2003).

35. Fioravante, D.; Regehr, W.G. Short-term forms of presynaptic plasticity. Current Opinion in Neurobiology, **21**, 269-274 (2011)

36. Atluri, P. P.; Regehr, W. G. Determinants of the Time Course of Facilitation at the Granule Cell to Purkinje Cell Synapse. J. Neurosci. **16**, 5661–5671 (1996)

37. Buonomano, D. V.; Maass, W. State-dependent computations: spatiotemporal processing in cortical networks. Nat. Rev. Neurosci. **10**, 113-698 (2009)

38. Fortune, E.S.; Rose, G.J. Short-term synaptic plasticity as a temporal filter Trends in Neurosciences, **24**, 381-385 (2001).

39. Abbott1,L.F.; Regehr, W.G. Synaptic computation. Nature. **431,** 796-803 (2004).

40. Yoneyama, M.; Fukushima, Y.; Tsukada, M.; Aihara, T. Spatiotemporal characteristics of synaptic EPSP summation on the dendritic trees of hippocampal CA1 pyramidal neurons as revealed by laser uncaging stimulation. Cogn Neurodyn. **5**, 333-342 (2011).